\documentclass{elsart}

\usepackage{amssymb}
\usepackage{graphicx}
\usepackage[ulem=normalem]{changes}

\begin{document}

\begin{frontmatter}

\title{Higher-rank discrete symmetries in the IBM.\\
III Tetrahedral shapes}

\author[ganil]{P.~Van~Isacker},
\author[batna]{A.~Bouldjedri}
and \author[batna]{S.~Zerguine}

\address[ganil]{Grand Acc\'el\'erateur National d'Ions Lourds, CEA/DRF--CNRS/IN2P3\\
Bvd Henri Becquerel, F-14076 Caen, France}

\address[batna]{Department of Physics, PRIMALAB Laboratory, Batna 1 University\\
Route de Biskra, 05000 Batna, Algeria}

\begin{abstract}
In the context of the \mbox{$sf$-IBM},
the interacting boson model with $s$ and $f$ bosons,
the conditions are derived for a rotationally invariant and parity-conserving Hamiltonian
with up to two-body interactions
to have a minimum with tetrahedral shape in its classical limit.
A degenerate minimum that includes a shape with tetrahedral symmetry
can be obtained in the classical limit of a Hamiltonian
that is transitional between the two limits of the model,
${\rm U}_f(7)$ and ${\rm SO}_{sf}(8)$.
The conditions for the existence of such a minimum are derived.
The system can be driven towards an isolated minimum with tetrahedral shape
through a modification of two-body interactions between the $f$ bosons.
General comments are made on the observational consequences
of the occurrence of shapes with a higher-rank discrete symmetry
in the context of algebraic models.
\end{abstract}

\begin{keyword}
discrete tetrahedral symmetry \sep
interacting boson model \sep
$f$ bosons

\PACS 21.60.Ev \sep 21.60.Fw 
\end{keyword}

\end{frontmatter}

\section{Introduction}
\label{s_intro}
This paper is a continuation of Refs.~\cite{Isacker15,Bouldjedri},
henceforth referred to as I and II,
as part of a series concerning nuclear shapes with a higher-rank discrete symmetry
in the framework of the interacting boson model (IBM)
and its possible extensions~\cite{Iachello87}.
In I and II we considered the case of hexadecapole deformation
giving rise to shapes with octahedral symmetry
and their manifestation in the \mbox{$sdg$-IBM}.
In the present paper we turn our attention to tetrahedral symmetry.

Shapes with tetrahedral discrete symmetry occur in lowest order
through a particular kind of octupole deformation,
namely  $Y_{3\mu}(\theta,\phi)$ with $\mu=\pm2$,
and all other deformations equal to zero~\cite{Li94,Dudek02,Dudek06}.
Whereas evidence for hexadecapole deformation in nuclei is circumstantial at best,
such is not the case for the octupole degree of freedom.
Octupole excitations in spherical nuclei are well documented
(see, e.g., the review~\cite{Butler96})
and there is even experimental evidence for nuclei
with a permanent octupole deformation~\cite{Gaffney13}.
This makes the search for nuclear shapes with tetrahedral symmetry all the more compelling.

The algebraic description of the octupole degree of freedom
requires the introduction of an $f$ boson
with angular momentum $\ell=3$ and negative parity,
as was already suggested in the early papers on the IBM~\cite{Arima76,Arima78,Scholten78}.
In principle, the $f$ boson should be considered in addition
to the bosons of the elementary version of the model
since for a realistic description of nuclear collective behavior
the quadrupole degree of freedom, and therefore the $d$ boson, cannot be neglected.
Furthermore, an octupole deformation causes a shift in the center of mass
that must be balanced by a dipole deformation,
which necessitates the introduction of a $p$ boson~\cite{Engel85}.
One concludes therefore that the search for tetrahedral deformation
should be carried out in the framework of the \mbox{$spdf$-IBM},
the properties of which have been studied in detail
in Refs.~\cite{Engel87,Kusnezov89,Kusnezov90}.
Unfortunately, a catastrophe analysis of this model
is a rather complicated problem
and the following simplification suggests itself
based on our experience with the search for octahedral deformation
in the context of the \mbox{$sdg$-IBM}.
Because quadrupole deformations must vanish
for the nucleus to acquire a shape with a higher-rank discrete symmetry,
it transpires that the $d$ boson is not an essential ingredient in our search,
the reason being that it should not or only weakly couple to the other bosons.
In fact, the most important conditions for the realization
of a shape with octahedral symmetry, as obtained in the \mbox{$sdg$-IBM} in I and II,
could just as well have been derived in the context of the \mbox{$sg$-IBM}.
By analogy, we suggest therefore that a search for tetrahedral deformation
in an algebraic context can be carried out in the simpler \mbox{$sf$-IBM},
which is the subject matter of the present paper.
It should be recognized however
that the absence of a rotational SU(3) limit in the \mbox{$sf$-IBM}
constitutes a limitation of the present approach.

The paper is structured as follows.
In Section~\ref{s_shape} we recall the parameterization of octupole shapes
and how, within this parameterization, a shape with tetrahedral symmetry can be realized.
Section~\ref{s_sfibm} introduces
the rotationally invariant, parity-conserving Hamiltonian of the \mbox{$sf$-IBM}
with up to two-body interactions,
of which the dynamical symmetries are discussed in Section~\ref{s_limits}
and the classical limit in Section~\ref{s_clas}.
The main results of this paper are presented in Section~\ref{s_tetra},
where a catastrophe analysis of the classical energy surface is carried out
to unveil the existence of minima at shapes with tetrahedral symmetry.
Finally, in Section~\ref{s_conc} the conclusions of this work are summarized.

\section{Octupole and tetrahedral shapes}
\label{s_shape}
In case of a pure octupole deformation 
seven variables $\alpha_{3\mu}$ are needed
to define the intrinsic shape
as well as the orientation of that shape in the laboratory frame.
One is therefore confronted with the problem
of the separation of intrinsic from orientation variables.
While this problem has a natural solution
in the case of quadrupole deformation~\cite{Bohr52,Bohr53,BM75},
namely intrinsic axes that are defined
by the mutually perpendicular symmetry planes of the quadrupole shape,
no such solution presents itself
in the case of octupole deformation~\cite{Rohozinski88}.
The parameterization of Hamamoto {\it et al.}~\cite{Hamamoto91} is used in the following
and the surface is written as
\begin{eqnarray}
R_{\rm o}(\theta,\phi)&=&
R_0\Biggl[1+
a_{30}Y_{30}(\theta,\phi)
\nonumber\\&&\qquad\;
+\sum_{\mu=1}^3a_{3\mu}Y_{3\mu}^{\pi_\mu}(\theta,\phi)-
\imath\sum_{\mu=1}^3b_{3\mu}Y_{3\mu}^{-\pi_\mu}(\theta,\phi)\Biggr],
\label{e_surface}
\end{eqnarray}
with $\pi_\mu\equiv(-)^\mu$ and where the combinations
\begin{equation}
Y_{\lambda\mu}^\pm(\theta,\phi)=
\frac{1}{\sqrt2}\left[Y_{\lambda\mu}(\theta,\phi)\pm Y_{\lambda-\mu}(\theta,\phi)\right],
\label{e_ylm}
\end{equation}
are introduced in terms of the usual spherical harmonics $Y_{\lambda\mu}(\theta,\phi)$.
The surface $R_{\rm o}(\theta,\phi)$ is determined by the seven (real) variables
$\{a_{30},a_{3\mu},b_{3\mu},\mu=1,2,3\}$.
Hamamoto {\it et al.}~\cite{Hamamoto91} define the intrinsic shape
through the four variables $\{\beta_3,\delta_3,\vartheta_3,\varphi_3\}$
\begin{eqnarray}
b_{32}&=&\beta_3\sin\delta_3,
\nonumber\\
a_{30}&=&\beta_3\cos\delta_3\sin\vartheta_3\cos\varphi_3,
\nonumber\\
\sqrt{\textstyle{\frac{3}{8}}}a_{31}-\sqrt{\textstyle{\frac{5}{8}}}a_{33}&=&\beta_3\cos\delta_3\sin\vartheta_3\sin\varphi_3,
\nonumber\\
\sqrt{\textstyle{\frac{3}{8}}}b_{31}+\sqrt{\textstyle{\frac{5}{8}}}b_{33}&=&\beta_3\cos\delta_3\cos\vartheta_3,
\label{e_param}
\end{eqnarray}
while three combinations are set to zero,
\begin{equation}
a_{32}=
\sqrt{\textstyle{\frac58}}a_{31}+\sqrt{\textstyle{\frac38}}a_{33}=
-\sqrt{\textstyle{\frac58}}b_{31}+\sqrt{\textstyle{\frac38}}b_{33}=0.
\label{e_constraints}
\end{equation}
All possible intrinsic octupole-deformed shapes are covered
by the following three ranges of parameters:
\begin{eqnarray}
{\rm (a)}&&\beta_3>0,
\quad
-\textstyle{\frac12}\pi<\delta_3<\textstyle{\frac12}\pi,
\quad
\tan^{-1}\sqrt{2}\leq\vartheta_3<\textstyle{\frac12}\pi,
\quad
0<\varphi_3\leq\textstyle{\frac14}\pi,
\nonumber\\
{\rm (b)}&&\beta_3>0,
\quad{\rm if}\quad
\delta_3=\textstyle{\frac12}\pi,
\nonumber\\
{\rm (c)}&&\beta_3>0,
\quad
0\leq\delta_3<\textstyle{\frac12}\pi,
\quad
0\leq\varphi_3\leq\textstyle{\frac14}\pi,
\quad{\rm if}\quad
\vartheta_3=\textstyle{\frac12}\pi,
\label{e_range}
\end{eqnarray}
where for range (a) the additional constraint $(\tan\vartheta_3)(\sin\varphi_3)\geq1$
should be satisfied.
The parameterization~(\ref{e_range}) has the important property
that a given intrinsic shape occurs only once over the entire range.

A shape with tetrahedral symmetry implies a vanishing quadrupole deformation, $\beta_2=0$,
and can be realized in lowest order with an octupole deformation
with $\mu=\pm2$~\cite{Dudek03,Dudek10}.
For the octupole parameterization~(\ref{e_param})
this implies $\beta_3>0$ and $\delta_3={\frac12}\pi$,
in which case the nuclear surface~(\ref{e_surface}) reduces to
\begin{equation}
\frac{R_{\rm o}(\theta,\phi)}{R_0}=
1+\imath\beta_3Y_{32}^-(\theta,\phi)=
1-\sqrt{\frac{105}{16\pi}}\beta_3(\sin\theta)^2\cos\theta\sin2\phi.
\label{e_surface1a}
\end{equation}
A single parameter, $\beta_3$, defines the surface with tetrahedral symmetry.

\section{The $sf$ interacting boson model}
\label{s_sfibm}
In this section the most general rotationally invariant and parity-conserving \mbox{$sf$-IBM} Hamiltonian
with up to two-body interactions is presented.
It has the same formal expression as given in I
with the additional constraint that parity is conserved.

A Hamiltonian of the \mbox{$sf$-IBM} conserves the total number of bosons
and can therefore be written in terms of the $(1+7)^2=64$ operators $b_{\ell m}^\dag b_{\ell' m'}$,
where $b_{\ell m}^\dag$ ($b_{\ell m}$) creates (annihilates)
a boson with angular momentum $\ell$ and $z$ projection $m$.
A boson-number-conserving Hamiltonian with up to two-body interactions is of the form
\begin{equation}
\hat H=\hat H_1+\hat H_2,
\label{e_ham}
\end{equation}
with a one-body term
\begin{eqnarray}
\hat H_1&=&
\epsilon_s[s^\dag\times\tilde s]^{(0)}-
\epsilon_f\sqrt{7}[f^\dag\times\tilde f]^{(0)}
\nonumber\\&=&
\epsilon_s\,s^\dag\cdot\tilde s+
\epsilon_f\,f^\dag\cdot\tilde f=
\epsilon_s\hat n_s+
\epsilon_f\hat n_f,
\label{e_ham1}
\end{eqnarray}
and a two-body interaction
\begin{equation}
\hat H_2=
\sum_{\ell_1\leq\ell_2,\ell'_1\leq\ell'_2,L}
\frac{(-)^Lv^L_{\ell_1\ell_2\ell'_1\ell'_2}}{\sqrt{(1+\delta_{\ell_1\ell_2})(1+\delta_{\ell'_1\ell'_2})}}
[b^\dag_{\ell_1}\times b^\dag_{\ell_2}]^{(L)}\cdot
[\tilde b_{\ell'_2}\times\tilde b_{\ell'_1}]^{(L)},
\label{e_ham2}
\end{equation}
with $\tilde b_{\ell m}\equiv(-)^{\ell-m}b_{\ell,-m}$.
The multiplication $\times$ refers to coupling in angular momentum
(shown as an upper-index in round brackets),
the dot $\cdot$ indicates a scalar product,
$b^\dag_\ell\cdot\tilde b_\ell\equiv\sum_mb^\dag_{\ell m}b_{\ell m}$,
$\hat n_\ell$ is the number operator for the $\ell$ boson
and the coefficient $\epsilon_\ell$ is its energy.
The coefficients $v^L_{\ell_1\ell_2\ell'_1\ell'_2}$ are the interaction matrix elements
between normalized two-boson states, 
$v^L_{\ell_1\ell_2\ell'_1\ell'_2}\equiv
\langle\ell_1\ell_2;LM_L|\hat H_2|\ell'_1\ell'_2;LM_L\rangle$.
Conservation of parity implies that this interaction matrix element vanishes
unless $(-)^{\ell_1+\ell_2}=(-)^{\ell'_1+\ell'_2}$.
Also, it will be assumed in the following that all Hamiltonians are Hermitian
so that $v^L_{\ell_1\ell_2\ell'_1\ell'_2}=v^L_{\ell'_1\ell'_2\ell_1\ell_2}$.

\section{Dynamical symmetries of the \mbox{$sf$-IBM}}
\label{s_limits}
Although the \mbox{$sf$-IBM} is a schematic model,
it of some interest to study its dynamical symmetries
since these correspond to two possible, basic manifestations
of octupole collectivity in nuclei.

The 64 operators $b_{\ell m}^\dag b_{\ell' m'}$ with $\ell,\ell'=0,3$
generate the Lie algebra U(8)
whose substructure therefore determines
the dynamical symmetries of the \mbox{$sf$-IBM}.
The first limit is obtained by eliminating from the generators of U(8)
those that involve the $s$ boson;
it is specified by the following chain of nested algebras:
\begin{equation}
({\rm I})\qquad
\begin{array}{ccccccc}
{\rm U}_{sf}(8)&\supset&{\rm U}_f(7)&\supset&{\rm SO}_f(7)&\supset&{\rm SO}_f(3)\\
\downarrow&&\downarrow&&\downarrow&&\downarrow\\[0mm]
[N]&&n_f&&\upsilon_f&\nu_f&L
\end{array},
\label{e_u7}
\end{equation}
where the subscripts `$s$' and/or `$f$'
are a reminder of the bosons that make up the generators of the algebra (see below).
Below each algebra the associated quantum number is given:
$N$ is the total number of bosons,
$n_f$ is the number of $f$ bosons,
$\upsilon_f$ is the $f$-boson seniority
(i.e., the number of $f$ bosons not in pairs coupled to angular momentum zero)
and $L$ is the angular momentum generated by the $f$ bosons.
(Since $L$ coincides with the total angular momentum,
its subscript `$f$' is suppressed.)
Additional multiplicity labels, collectively denoted as $\nu_f$ and not associated to an algebra,
are needed between ${\rm SO}_f(7)$ and ${\rm SO}_f(3)$. 
In this limit, which shall be referred to as ${\rm U}_f(7)$ or limit I,
 the separate numbers of $s$ and $f$ bosons are conserved,
giving rise to a vibrational-like spectrum
with a spherical shape of the ground state
and oscillations in the octupole degree of freedom.

The second dynamical symmetry corresponds to the following chain of nested algebras:
\begin{equation}
({\rm II})\qquad
\begin{array}{ccccccc}
{\rm U}_{sf}(8)&\supset&{\rm SO}_{sf}(8)&\supset&{\rm SO}_f(7)&\supset&{\rm SO}_f(3)\\
\downarrow&&\downarrow&&\downarrow&&\downarrow\\[0mm]
[N]&&\upsilon_{sf}&&\upsilon_f&\nu_f&L
\end{array}.
\label{e_so8}
\end{equation}
The algebras and quantum numbers are identical to those in the vibrational limit~(\ref{e_u7})
but for the appearance of ${\rm SO}_{sf}(8)$ and its associated label $\upsilon_{sf}$,
resulting from the pairing of $s$ and $f$ bosons.
As shown in Section~\ref{s_clas},
the ground state acquires a permanent octupole deformation in this limit,
which shall be referred to as ${\rm SO}_{sf}(8)$ or limit II.

The dynamical symmetries of ${\rm U}_{sf}(8)$
describe the two basic manifestations of octupole collectivity in nuclei:
octupole vibrations around a spherical shape (limit I)
or a permanent octupole deformation (limit II).
The latter limit is of relevance in the search for tetrahedral deformation
but it has the unrealistic feature
that the energies of the $s$ and $f$ boson are taken to be degenerate.
In Sections~\ref{s_clas} and~\ref{s_tetra}
we investigate to what extent non-degenerate single-boson energies can be accommodated 
while still preserving an octupole-deformed minimum,
and whether that minimum can have tetrahedral symmetry.

For further reference, we list some of the properties of limits I and II.
The classification of limits I and II can be summarized with the algebraic lattice
\begin{equation}
\begin{array}{c}
{\rm U}_{sf}(8)\\
\swarrow\quad\searrow\\
\quad {\rm U}_f(7)\quad {\rm SO}_{sf}(8)\\
\searrow\quad\swarrow\\
{\rm SO}_f(7)\\
\downarrow\\
{\rm SO}_f(3)
\end{array}.
\label{e_lat}
\end{equation}
The generators of the different subalgebras in the lattice~(\ref{e_lat}) are
\begin{eqnarray}
{\rm U}_f(7)&:\quad&
\{[f^\dag\times\tilde f]^{(\lambda)}_\mu,\lambda=0,\dots,6\},
\nonumber\\
{\rm SO}_{sf}(8)&:\quad&
\{[s^\dag\times\tilde f-f^\dag\times\tilde s]^{(3)}_\mu,
[f^\dag\times\tilde f]^{(\lambda)}_\mu,\lambda=1,3,5\},
\nonumber\\
{\rm SO}_f(7)&:\quad&
\{[f^\dag\times\tilde f]^{(\lambda)}_\mu,\lambda=1,3,5\},
\nonumber\\
{\rm G}_2&:\quad&
\{[f^\dag\times\tilde f]^{(\lambda)}_\mu,\lambda=1,5\},
\nonumber\\
{\rm SO}_f(3)&:\quad&
\{\hat L_\mu\equiv\sqrt{28}[f^\dag\times\tilde f]^{(1)}_\mu\}.
\label{e_gens}
\end{eqnarray}
Note the presence of the additional (exceptional) algebra ${\rm G}_2$,
which occurs in between ${\rm SO}_f(7)$ and ${\rm SO}_f(3)$~\cite{Racah49}.
It does not appear in Eqs.~(\ref{e_u7}) and~(\ref{e_so8})
because in symmetric irreducible representations
the quadratic Casimir operators of ${\rm SO}_f(7)$ and ${\rm G}_2$
have identical expectation values.
The exceptional algebra ${\rm G}_2$ is therefore
discarded from the classifications~(\ref{e_u7}), (\ref{e_so8}) and (\ref{e_lat})
without loss of generality.

The explicit expressions of linear and quadratic Casimir operators
of the algebras appearing in the lattice~(\ref{e_lat}) are
\begin{eqnarray}
\hat C_1[{\rm U}_{sf}(8)]&=&\hat N=\hat n_s+\hat n_f,
\nonumber\\
\hat C_2[{\rm U}_{sf}(8)]&=&\hat N(\hat N+7),
\nonumber\\
\hat C_1[{\rm U}_f(7)]&=&\hat n_f,
\nonumber\\
\hat C_2[{\rm U}_f(7)]&=&\hat n_f(\hat n_f+6),
\nonumber\\
\hat C_2[{\rm SO}_{sf}(8)]&=&
[s^\dag\times\tilde f-f^\dag\times\tilde s]^{(3)}\cdot[s^\dag\times\tilde f-f^\dag\times\tilde s]^{(3)}+
\hat C_2[{\rm SO}_f(7)],
\nonumber\\
\hat C_2[{\rm SO}_f(7)]&=&
2\sum_{\lambda\;{\rm odd}}[f^\dag\times\tilde f]^{(\lambda)}\cdot[f^\dag\times\tilde f]^{(\lambda)},
\nonumber\\
\hat C_2[{\rm SO}_f(3)]&=&\hat L\cdot\hat L.
\label{e_cas}
\end{eqnarray}
The expressions for the quadratic Casimir operators
$\hat C_2[{\rm U}_{sf}(8)]$ and $\hat C_2[{\rm U}_f(7)]$
are not general but are valid in symmetric irreducible representations of
${\rm U}_{sf}(8)$ and ${\rm U}_f(7)$.
A rotationally invariant and parity-conserving Hamiltonian with up to two-body interactions
can be written in terms of the Casimir operators~(\ref{e_cas}),
\begin{eqnarray}
\hat H_{\rm sym}&=&
\epsilon_s\,\hat n_s+\epsilon_f\,\hat n_f+
a_f\,\hat C_2[{\rm U}_f(7)]+
b_{sf}\,\hat P_{sf}^\dag\hat P_{sf}+
b_f\,\hat C_2[{\rm SO}_f(7)]
\nonumber\\&&+
c_f\,\hat C_2[{\rm SO}_f(3)],
\label{e_hamlat}
\end{eqnarray}
where $\epsilon_\ell$, $a_\ell$, $b_\ell$, $b_{\ell\ell'}$ and $c_\ell$ are parameters.
The quadratic Casimir operator of ${\rm U}_{sf}(8)$ is omitted for simplicity
since it gives a constant contribution for a fixed boson number $N=n_s+n_f$.
The pairing interaction for $s$ and $f$ bosons
can be expressed in terms of Casimir operators,
\begin{equation}
\hat P^\dag_{sf}\hat P_{sf}=
\hat C_2[{\rm U}_{sf}(8)]-\hat C_1[{\rm U}_{sf}(8)]-\hat C_2[{\rm SO}_{sf}(8)],
\label{e_pairing}
\end{equation}
where $\hat P^\dag_{sf}\equiv s^\dag s^\dag-f^\dag\cdot f^\dag$.
Equation~(\ref{e_hamlat}) is the most general Hamiltonian with up to two-body interactions 
that can be written in terms of invariant operators of the lattice~(\ref{e_lat}).
It is intermediate between the limits I and II
but has less parameters than the general Hamiltonian~(\ref{e_ham}).
The latter contains seven boson--boson interaction matrix elements
whereas the symmetry Hamiltonian~(\ref{e_hamlat}) has only four two-body parameters.

The ${\rm U}_f(7)$ limit is attained for $b_{sf}=0$
leading to the eigenvalues 
\begin{equation}
E_{\rm I}=
\epsilon_s\,n_s+\epsilon_f\,n_f+
a_f\,n_f(n_f+6)+
b_f\,\upsilon_f(\upsilon_f+5)+
c_f\,L(L+1).
\label{e_eigu7}
\end{equation}
The ${\rm SO}_{sf}(8)$ limit occurs
for $\epsilon_s=\epsilon_f\equiv\epsilon_{sf}$ and $a_f=0$,
in which case the Hamiltonian's eigenstates have the eigenvalues
\begin{eqnarray}
E_{\rm II}&=&
\epsilon_{sf}\,N+
b_{sf}[N(N+6)-\upsilon_{sf}(\upsilon_{sf}+6)]+
b_f\,\upsilon_f(\upsilon_f+5)
\nonumber\\&&+
c_f\,L(L+1).
\label{e_eigso8}
\end{eqnarray}
The eigenspectra in two limits are then determined
with the help of the branching rules
\begin{eqnarray}
{\rm U}_{sf}(8)\supset{\rm U}_f(7)&\quad:\quad&[N]\mapsto n_f=0,1,\dots,N,
\nonumber\\
{\rm U}_f(7)\supset{\rm SO}_f(7)&\quad:\quad&n_f\mapsto\upsilon_f=n_f,n_f-2,\dots,1\;{\rm or}\;0,
\nonumber\\
{\rm U}_{sf}(8)\supset{\rm SO}_{sf}(8)&\quad:\quad&[N]\mapsto\upsilon_{sf}=N,N-2,\dots,1\;{\rm or}\;0,
\nonumber\\
{\rm SO}_{sf}(8)\supset{\rm SO}_f(7)&\quad:\quad&\upsilon_{sf}\mapsto\upsilon_f=0,1,\dots,\upsilon_{sf}.
\nonumber\\
\label{e_branch}
\end{eqnarray}
The ${\rm SO}_f(7)\supset{\rm SO}_f(3)$ reduction from seniority to angular momentum
is more complicated due to the multiplicity problem.
A closed formula is available
for the number of times the angular momentum $L$ occurs for a given seniority $\upsilon_f$
in terms of an integral over characters of the orthogonal algebras SO(7) and SO(3)~\cite{Weyl39}.
This number $d(\upsilon_f,L)$ is given by complex integral~\cite{Gheorghe04}
\begin{equation}
d(\upsilon_f,L)=
\frac{i}{2\pi}
\oint_{|z|=1}
\frac{(z^{2L+1}-1)(z^{2\upsilon_f+5}-1)\prod_{k=1}^4(z^{\upsilon_f+k}-1)}
{z^{3\upsilon_f+L+2}\prod_{k=1}^4(z^{k+1}-1)}dz,
\label{e_multiplicity}
\end{equation}
which, due to Cauchy's theorem, can be evaluated by taking the negative of the residue of its integrand.
An alternative recursive method to determine the ${\rm SO}_f(7)\supset{\rm SO}_f(3)$ reduction
was proposed by Rohozi\'nski~\cite{Rohozinski78}.
Tables of multiplicities $d(\upsilon_f,L)$ can be found in Refs.~\cite{Rohozinski78,Isacker14}.

\begin{figure}
\centering
\includegraphics[width=6.5cm]{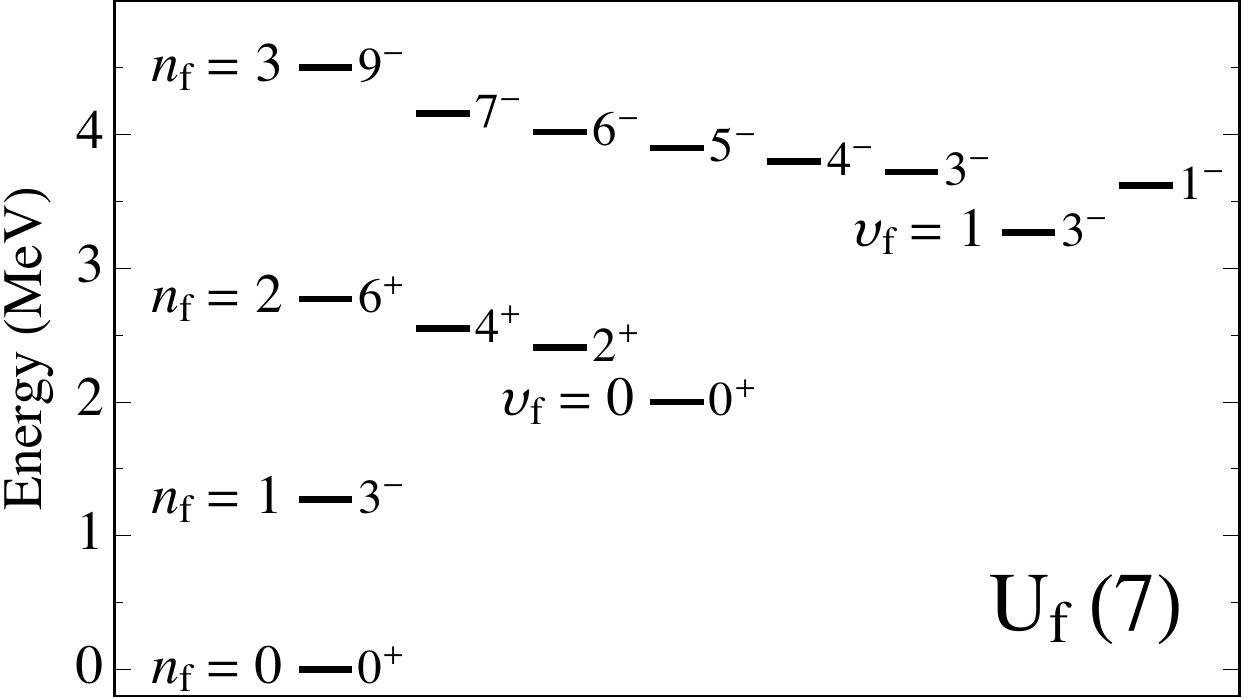}
\includegraphics[width=6.5cm]{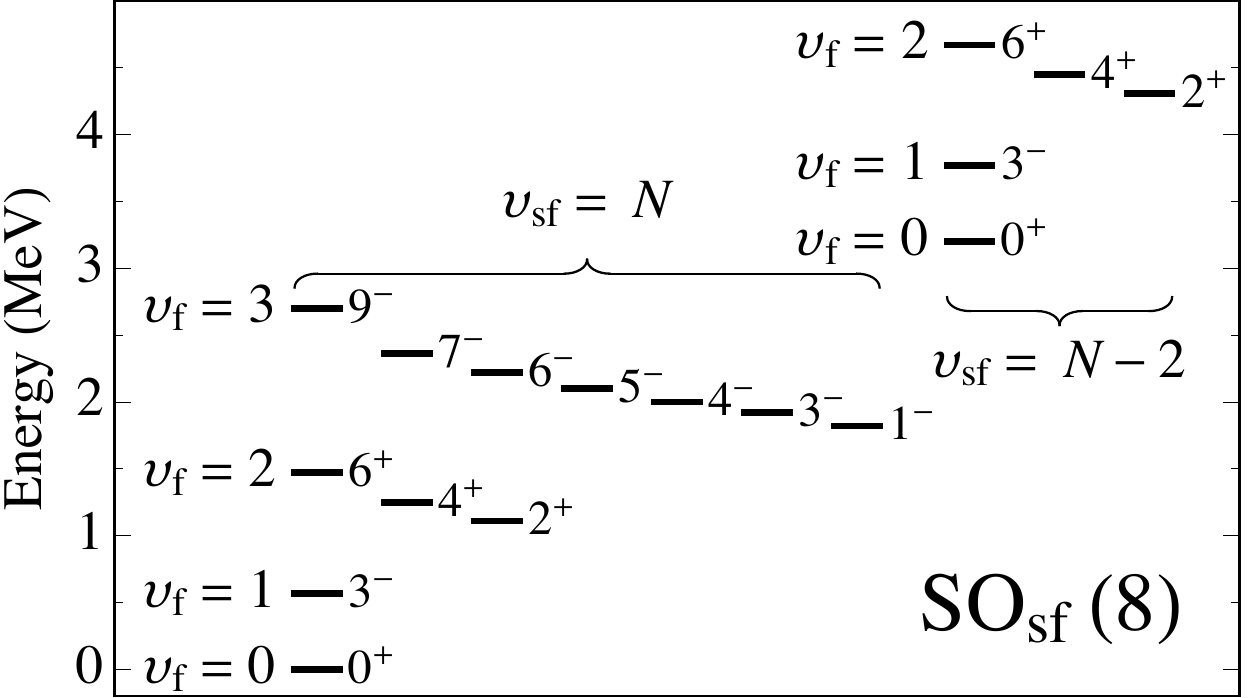}
\caption{Energy spectra in the ${\rm U}_f(7)$ and ${\rm SO}_{sf}(8)$ limits
of the \mbox{$sf$-IBM} for $N=6$ bosons.
For the ${\rm U}_f(7)$ spectrum the non-zero parameters in the Hamiltonian~(\ref{e_hamlat})
are $\epsilon_f-\epsilon_s=1000$, $b_f=25$ and $c_f=10$~keV.
For the ${\rm SO}_{sf}(8)$ spectrum the non-zero parameters
are $\epsilon_f-\epsilon_s=0$, $b_{sf}=100$, $b_f=75$ and $c_f=10$~keV.}
\label{f_limits}
\end{figure}
Typical energy spectra in the ${\rm U}_f(7)$ and ${\rm SO}_{sf}(8)$ limits are shown in Fig.~\ref{f_limits}.
The ${\rm U}_f(7)$ spectrum displays octupole-phonon multiplets
characterized by a fixed number of $f$ bosons, $n_f=0,1,\dots$
The multiplets are further structured by the seniority quantum number:
the $n_f=2$ multiplet has $\upsilon_f=2$ except for the $0^+$ level, which has $\upsilon_f=0$,
the $n_f=3$ multiplet has $\upsilon_f=3$ except for the $3^-$ level, which has $\upsilon_f=1$, etc.
The ${\rm SO}_{sf}(8)$ spectrum contains sets of levels with $\upsilon_{sf}=N,N-2,\dots$
and, due the repulsive $sf$-pairing, $\upsilon_{sf}=N$ levels are lowest in energy.
Multiplets characterized by a seniority quantum number $\upsilon_f=0,1,\dots$
occur within each ${\rm SO}_{sf}(8)$ multiplet.

\section{Classical limit of the \mbox{$sf$-IBM}}
\label{s_clas}
The classical limit of an arbitrary interacting boson Hamiltonian
is its expectation value in a coherent state~\cite{Gilmore79},
which is a function of the deformation variables
and is to be interpreted as a total-energy surface.
The method was first proposed for the \mbox{$sd$-IBM}~\cite{Ginocchio80,Dieperink80}.
The coherent state for the \mbox{$sf$-IBM} is inspired by the surface~(\ref{e_surface}),
\begin{equation}
|N;a_{3\mu},b_{3\mu}\rangle\propto
\Gamma(a_{3\mu},b_{3\mu})^N|{\rm o}\rangle,
\label{e_coh1a}
\end{equation}
with~\cite{note1}
\begin{equation}
\Gamma(a_{3\mu},b_{3\mu})=
s^\dag+a_{30}f^\dag_0+
\sum_{\mu=1}^3a_{3\mu}(f^{\pi_\mu}_\mu)^\dag+
\imath\sum_{\mu=1}^3b_{3\mu}(f^{-\pi_\mu}_\mu)^\dag,
\label{e_coh1b}
\end{equation}
where $|{\rm o}\rangle$ is the boson vacuum
and the creation operators are defined as
\begin{equation}
(f^\pm_\mu)^\dag=
\frac{1}{\sqrt2}\left(f_\mu^\dag\pm f_{-\mu}^\dag\right).
\label{e_foperator}
\end{equation}
The coefficients $a_{3\mu}$ and $b_{3\mu}$ have the interpretation of the shape variables
appearing in the expansion~(\ref{e_surface}).
In contrast to the geometric model of Bohr and Mottelson~\cite{BM75}
where deformation is associated with the entire nucleus,
in the IBM it is generated by the valence nucleons only.
As a result, the shape variables in both models
are proportional but not identical~\cite{Ginocchio80b}.
In the parameterization~(\ref{e_param})
the radial parameter $\beta_3$
in the geometric model and in the IBM are proportional
while the angles parameters have an identical interpretation.

The coherent state based on the parameterization~(\ref{e_param}) reads
\begin{equation}
|N;\beta_3,\delta_3,\vartheta_3,\varphi_3\rangle=
\sqrt{\frac{1}{N!(1+\beta_3^2)^N}}
\Gamma(\beta_3,\delta_3,\vartheta_3,\varphi_3)^N|{\rm o}\rangle,
\label{e_coh2a}
\end{equation}
with
\begin{eqnarray}
&&\Gamma(\beta_3,\delta_3,\vartheta_3,\varphi_3)
\nonumber\\&&=
s^\dag+
\beta_3\Bigl[\cos\delta_3\sin\vartheta_3\cos\varphi_3f^\dag_0+
\imath\sqrt{\textstyle{\frac12}}\sin\delta_3(f^\dag_{-2}-f^\dag_{+2})
\nonumber\\&&\qquad\quad
-\sqrt{\textstyle{\frac{3}{16}}}\cos\delta_3
\Bigl(\sin\vartheta_3\sin\varphi_3(f^\dag_{-1}-f^\dag_{+1})+
\imath\cos\vartheta_3(f^\dag_{-1}+f^\dag_{+1})\Bigr)
\nonumber\\&&\qquad\quad
+\sqrt{\textstyle{\frac{5}{16}}}\cos\delta_3
\Bigl(\sin\vartheta_3\sin\varphi_3(f^\dag_{-3}-f^\dag_{+3})-
\imath\cos\vartheta_3(f^\dag_{-3}+f^\dag_{+3})\Bigr)
\Bigr].
\label{e_coh2b}
\end{eqnarray}

The classical limit of a Hamiltonian of the \mbox{$sf$-IBM}
is its expectation value in the coherent state,
\begin{equation}
\langle\hat H\rangle\equiv
\langle N;\beta_3,\delta_3,\vartheta_3,\varphi_3|\hat H
|N;\beta_3,\delta_3,\vartheta_3,\varphi_3\rangle,
\label{e_clim1}
\end{equation}
which can be obtained by differentiation~\cite{Isacker81}.
The classical limit of the one-body part~(\ref{e_ham1}) is
\begin{equation}
\langle\hat H_1\rangle=
N\frac{\epsilon_s+\epsilon_f\beta_3^2}{1+\beta_3^2},
\label{e_clim11}
\end{equation}
and that of the two-body part~(\ref{e_ham2})
can be written in the generic form
\begin{equation}
\langle\hat H_2\rangle=
\frac{N(N-1)}{(1+\beta_3^2)^2}
\left[\sum_{l=0,2,4}c_l(\beta_3)^l+\Phi(\delta_3,\vartheta_3,\varphi_3)(\beta_3)^4\right],
\label{e_clim12}
\end{equation}
where
\begin{equation}
\Phi(\delta_3,\vartheta_3,\varphi_3)=
\sum_{ijk}(c_{ijk}+b_{ijk}\sin\delta_3\sin\varphi_3)(\cos\delta_3)^i(\cos\vartheta_3)^j(\cos\varphi_3)^k,
\label{e_phi}
\end{equation}
with coefficients $c_l$, $c_{ijk}$ and $b_{ijk}$
that can be expressed in terms of the interactions $v^L_{\ell_1\ell_2\ell'_1\ell'_2}$.
The expressions for the coefficients $c_l$ are
\begin{eqnarray}&&
\textstyle
c_0={\frac12}v_{ssss}^0,
\quad
c_2=v_{sfsf}^3-\sqrt{\frac17}v_{ssff}^0,
\nonumber\\&&
\textstyle
c_4=\frac{1}{14}v_{ffff}^0+\frac{3}{11}v_{ffff}^4+\frac{12}{77}v_{ffff}^6,
\label{e_coef1}
\end{eqnarray}
and those for the non-zero coefficients $c_{ijk}$ and $b_{ijk}$ are
\begin{eqnarray}&&
\textstyle
c_{200}=\frac{10}{231}\bar v,
\quad
c_{400}=-\frac{8}{231}\bar v,
\quad
c_{422}=-c_{424}=\frac{15}{308}\bar v,
\nonumber\\&&
\textstyle
c_{420}=c_{402}=-c_{440}=-c_{404}=c_{442}=-c_{444}=-\frac{15}{616}\bar v,
\nonumber\\&&
\textstyle
b_{311}=-b_{331}=\frac{\sqrt{15}}{77}\bar v,
\label{e_coef2}
\end{eqnarray}
in terms of the linear combination
\begin{equation}
\bar v\equiv11v_{ffff}^2-18v_{ffff}^4+7v_{ffff}^6.
\label{e_coef3}
\end{equation}

The classical limit of the total Hamiltonian~(\ref{e_ham})
can therefore be written as
\begin{eqnarray}
\langle\hat H\rangle&\equiv&
E(\beta_3,\delta_3,\vartheta_3,\varphi_3)
\nonumber\\&=&
\frac{N(N-1)}{(1+\beta_3^2)^2}
\left[\sum_{l=0,2,4}c'_l(\beta_3)^l+\Phi(\delta_3,\vartheta_3,\varphi_3)(\beta_3)^4\right],
\label{e_clim2}
\end{eqnarray}
where $c'_l$ are the modified coefficients
\begin{equation}
c'_0=c_0+\epsilon'_s,
\quad
c'_2=c_2+\epsilon'_s+\epsilon'_f,
\quad
c'_4=c_4+\epsilon'_f,
\label{e_coef4}
\end{equation}
in terms of the scaled boson energies $\epsilon'_\ell\equiv\epsilon_\ell/(N-1)$.

The quantum-mechanical Hamiltonian~(\ref{e_ham}), if it is Hermitian,
depends on two single-boson energies $\epsilon_\ell$
and seven two-body interactions $v^L_{\ell_1\ell_2\ell'_1\ell'_2}$.
In the classical limit with the coherent state~(\ref{e_coh1a}),
the number of independent parameters
in the energy surface $E(\beta_3,\delta_3,\vartheta_3,\varphi_3)$ is reduced to four
[three coefficients $c'_l$ and the single combination $\bar v$,
which determines completely the function $\Phi(\delta_3,\vartheta_3,\varphi_3)$].

\section{Tetrahedral shapes in the $sf$-IBM}
\label{s_tetra}
The question treated in this section is:
What are the conditions on the interactions in the \mbox{$sf$-IBM}
for the energy surface
$E(\beta_3,\delta_3,\vartheta_3,\varphi_3)$ in Eq.~(\ref{e_clim2})
to have a minimum with tetrahedral symmetry?
Fortunately, a complete catastrophe analysis of the surface
is not needed to answer this question.

The conditions for $E(\beta_3,\delta_3,\vartheta_3,\varphi_3)$
to have an extremum at a point $p^*$ in the four-dimensional space
of variables $\{\beta_3,\delta_3,\vartheta_3,\varphi_3\}$ are
\begin{equation}
\left.\frac{\partial E}{\partial\beta_3}\right|_{p^*}=
\left.\frac{\partial E}{\partial\delta_3}\right|_{p^*}=
\left.\frac{\partial E}{\partial\vartheta_3}\right|_{p^*}=
\left.\frac{\partial E}{\partial\varphi_3}\right|_{p^*}=0,
\label{e_extr}
\end{equation}
where $p^*\equiv(\beta^*_3,\delta^*_3,\vartheta^*_3,\varphi^*_3)$
is a short-hand notation for a critical point.
A critical point with tetrahedral symmetry will be denoted as $t^*$,
which implies that $t^*$
satisfies $\beta^*_3>0$ and $\delta^*_3={\frac12}\pi$.
The conditions~(\ref{e_extr}) are necessary
for $E(\beta_3,\delta_3,\vartheta_3,\varphi_3)$ to have an {\em extremum} at $p^*$;
the conditions for a {\em minimum} require in addition
that the eigenvalues of the stability matrix
[{\it i.e.}, the partial derivatives of $E(\beta_3,\delta_3,\vartheta_3,\varphi_3)$ of second order at $p^*$]
are all non-negative.

Three out of the four conditions~(\ref{e_extr})
are always satisfied for $p^*=t^*$.
The fourth, namely the one related to the partial derivative in $\beta_3$,
leads to a cubic equation in $\beta_3$ with the solutions
\begin{equation}
\beta_3^*=0,
\qquad
\beta_3^*=\pm\sqrt{\frac{2c'_0-c'_2}{2c'_4-c'_2}}.
\label{e_extb3}
\end{equation}
Only the last solution with a plus sign corresponds to a tetrahedral extremum
and therefore the following condition on the ratio of coefficients is obtained:
\begin{equation}
\frac{2c'_0-c'_2}{2c'_4-c'_2}>0.
\label{e_ratio1}
\end{equation}
 
The partial derivatives of $E(\beta_3,\delta_3,\vartheta_3,\varphi_3)$ of second order
are identically zero at $p^*=t^*$,
except the double derivatives in $\beta_3$ and $\delta_3$.
For the eigenvalues of the stability matrix to be positive
the following two conditions must be satisfied:
\begin{equation}
\frac{(2c'_0-c'_2)(2c'_4-c'_2)^3}{(c'_0-c'_2+c'_4)^3}>0,
\qquad
\frac{(2c'_0-c'_2)^2c_{200}}{(c'_0-c'_2+c'_4)^2}>0.
\label{e_ratio2}
\end{equation}
The condition~(\ref{e_ratio1}) for an extremum with tetrahedral symmetry,
combined with the conditions~(\ref{e_ratio2}) that the extremum is a minimum,
therefore lead to
\begin{equation}
2c'_0-c'_2>0,
\qquad
2c'_4-c'_2>0,
\qquad
c_{200}>0,
\label{e_condi}
\end{equation}
which translate into the following conditions
on the single-boson energies and interaction matrix elements:
\begin{eqnarray}&&
\textstyle
(N-1)\left(v_{ssss}^0-v_{sfsf}^3+\sqrt{\frac17}v_{ssff}^0\right)>\epsilon_f-\epsilon_s,
\nonumber\\&&
\textstyle
(N-1)\left(\frac{1}{7}v_{ffff}^0+\frac{6}{11}v_{ffff}^4+\frac{24}{77}v_{ffff}^6
-v_{sfsf}^3+\sqrt{\frac17}v_{ssff}^0\right)>\epsilon_s-\epsilon_f,
\nonumber\\&&
\textstyle
11v_{ffff}^2-18v_{ffff}^4+7v_{ffff}^6>0.
\label{e_cond}
\end{eqnarray}
These are the necessary and sufficient conditions
for the general Hamiltonian of the \mbox{$sf$-IBM},
Eqs.~(\ref{e_ham}), (\ref{e_ham1}) and (\ref{e_ham2}),
to have a minimum with tetrahedral shape in its classical limit.

Can these conditions be fulfilled for ``realistic'' values of single-boson energies
and boson--boson interaction matrix elements?
To answer this question,
let us first consider the most general Hamiltonian of the \mbox{$sf$-IBM}
except for one matrix element, namely $v^0_{ssff}$,
which is assumed to be zero.
This Hamiltonian is not analytically solvable
but the energies of its $0^+$ ground state
and its yrast $3^-$ state are known in closed form:
\begin{eqnarray}
E(0^+_1)&=&\textstyle N\epsilon_s+{\frac12}N(N-1)v^0_{ssss},
\nonumber\\
E(3^-_1)&=&\textstyle(N-1)\epsilon_s+\epsilon_f+(N-1)v^3_{sfsf}+{\frac12}(N-1)(N-2)v^0_{ssss},
\label{e_vibcondi}
\end{eqnarray}
resulting in
\begin{equation}
E(3^-_1)-E(0^+_1)=\epsilon_f-\epsilon_s-(N-1)(v^0_{ssss}-v^3_{sfsf}).
\label{e_vibcond}
\end{equation}
Therefore, unless $v^0_{ssff}>0$, the first of the conditions~(\ref{e_cond}) implies
that $E(3^-_1)<E(0^+_1)$, which is clearly unphysical.

One concludes therefore that the minimum
in the energy surface $E(\beta_3,\delta_3,\vartheta_3,\varphi_3)$ in Eq.~(\ref{e_clim2})
can be of tetrahedral shape only if the mixing matrix element $v^0_{ssff}$ is non-zero.
This brings us to the study of the symmetry Hamiltonian~(\ref{e_hamlat}),
which has the classical limit
\begin{equation}
\langle\hat H_{\rm sym}\rangle=
N\frac{\epsilon_s+\Gamma_f\beta_3^2}{1+\beta_3^2}+
N(N-1)\left[\frac{a_f\beta_3^4}{(1+\beta_3^2)^2}+
b_{sf}\left(\frac{1-\beta_3^2}{1+\beta_3^2}\right)^2\right].
\label{e_claslat}
\end{equation}
where the combination of parameters
$\Gamma_f=\epsilon_f+7a_f+6b_f+12c_f$ is introduced.
The parameter $b_{sf}$ is the pairing strength
for $s$ and $f$ bosons and is positive,
such that the ground-state configuration has $\upsilon_{sf}=N$,
akin to the situation in the SO(6) limit of the \mbox{$sd$-IBM}~\cite{Arima79}.
Provided $b_{sf}$ is large enough,
the energy surface~(\ref{e_claslat}) has an octupole-deformed minimum
($\beta_3^*\approx1$ for $b_{sf}\rightarrow\infty$)
but the shape at minimum is pear-like and not tetrahedral.
It can be concluded therefore that no isolated minimum with tetrahedral symmetry
occurs in the classical limit of the symmetry Hamiltonian~(\ref{e_hamlat}).
What still can happen, however, is that
a degenerate minimum occurs with non-zero octupole deformation,
which, given the instability in $\delta_3$, {\em includes} a tetrahedral shape.

The fact that no isolated tetrahedral minimum occurs in the classical limit
of the symmetry Hamiltonian~(\ref{e_hamlat})
can be understood from the conditions~(\ref{e_cond}),
the first and second of which reduce to
\begin{eqnarray}
4b_{sf}(N-1)-7a_f-6b_f-12c_f&>&\epsilon_f-\epsilon_s,
\nonumber\\
4b_{sf}(N-1)+a_f(2N+5)+6b_f+12c_f&>&\epsilon_s-\epsilon_f.
\label{e_conlat12}
\end{eqnarray}
Both inequalities can be satisfied provided $b_{sf}$ is positive and large enough.
On the other hand, the last of the conditions~(\ref{e_cond}) is not satisfied
because the combination of $f$-boson two-body matrix elements
vanishes identically for the symmetry Hamiltonian~(\ref{e_hamlat}),
\begin{equation}
22(a_f+b_f-9c_f)-
36(a_f+b_f-2c_f)+
14(a_f+b_f+9c_f)=0.
\label{e_conlat3}
\end{equation}
The absence of a tetrahedral minimum for the symmetry Hamiltonian~(\ref{e_hamlat})
is therefore entirely due to the specific combination of $f$-boson two-body matrix elements,
of which nothing is known, either empirically or microscopically.
If $v_{ffff}^2$ is taken more repulsive,
the energy surface in the classical limit of the \mbox{$sf$-IBM} Hamiltonian
acquires a minimum with tetrahedral symmetry.
Indeed, this modification does not alter the conditions~(\ref{e_conlat12})
since the matrix element $v_{ffff}^2$ does not appear in them,
whereas the third of the conditions~(\ref{e_cond}) is now satisfied.
A possible procedure to construct a Hamiltonian in the \mbox{$sf$-IBM}
with a minimum with tetrahedral shape in its classical limit
is therefore to add to an octupole-deformed symmetry Hamiltonian~(\ref{e_hamlat})
a repulsive $v_{ffff}^2$ interaction.

\begin{figure}
\centering
\includegraphics[width=7cm]{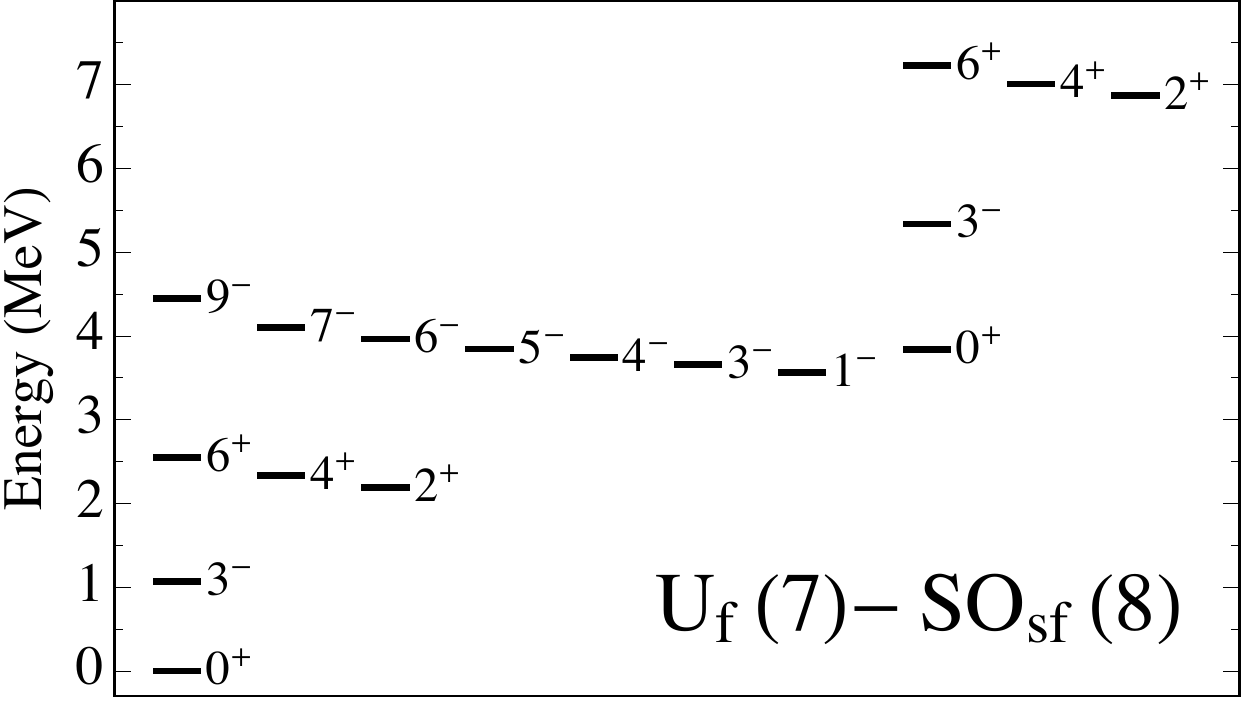}
\caption{Energy spectrum of a ${\rm U}_f(7)$--${\rm SO}_{sf}(8)$ transitional Hamiltonian
of the \mbox{$sf$-IBM} for $N=6$ bosons.
The non-zero parameters of the Hamiltonian~(\ref{e_hamlat})
are $\epsilon_f-\epsilon_s=1200$, $b_{sf}=100$, $b_f=50$ and $c_f=10$~keV.}
\label{f_u7so8}
\end{figure}
We illustrate this procedure with an example,
starting from a ${\rm U}_f(7)$--${\rm SO}_{sf}(8)$ transitional Hamiltonian
associated with the lattice~(\ref{e_lat}),
giving rise to the spectrum shown in Fig.~\ref{f_u7so8}.
A reasonable energy difference between the $s$ and $f$ bosons is taken
and the strength of the $sf$ pairing is chosen
so as to obtain an octupole-deformed minimum.
Other parameters in the Hamiltonian~(\ref{e_hamlat}) are of lesser importance.

\begin{figure}
\centering
\includegraphics[width=4cm]{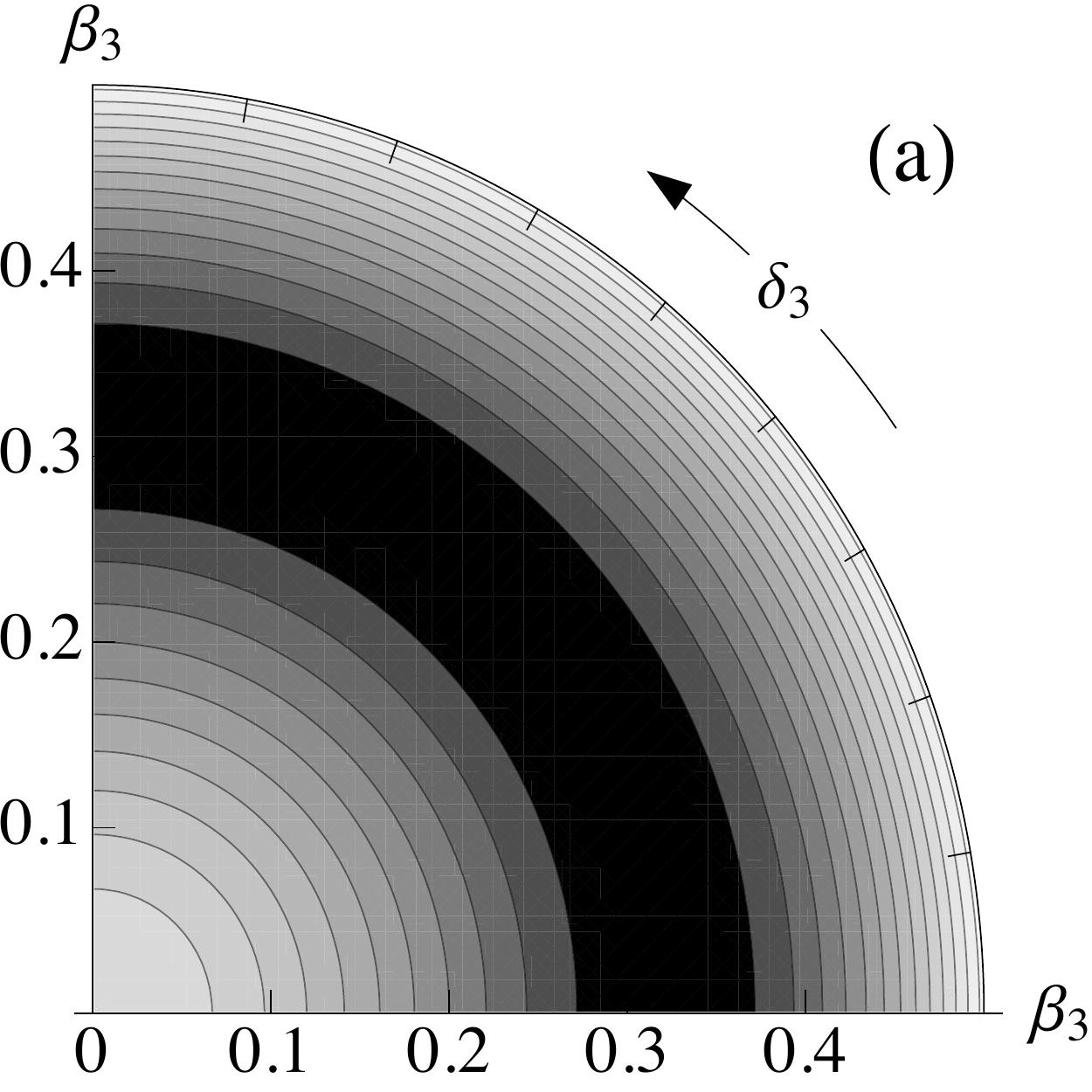}
\includegraphics[width=4cm]{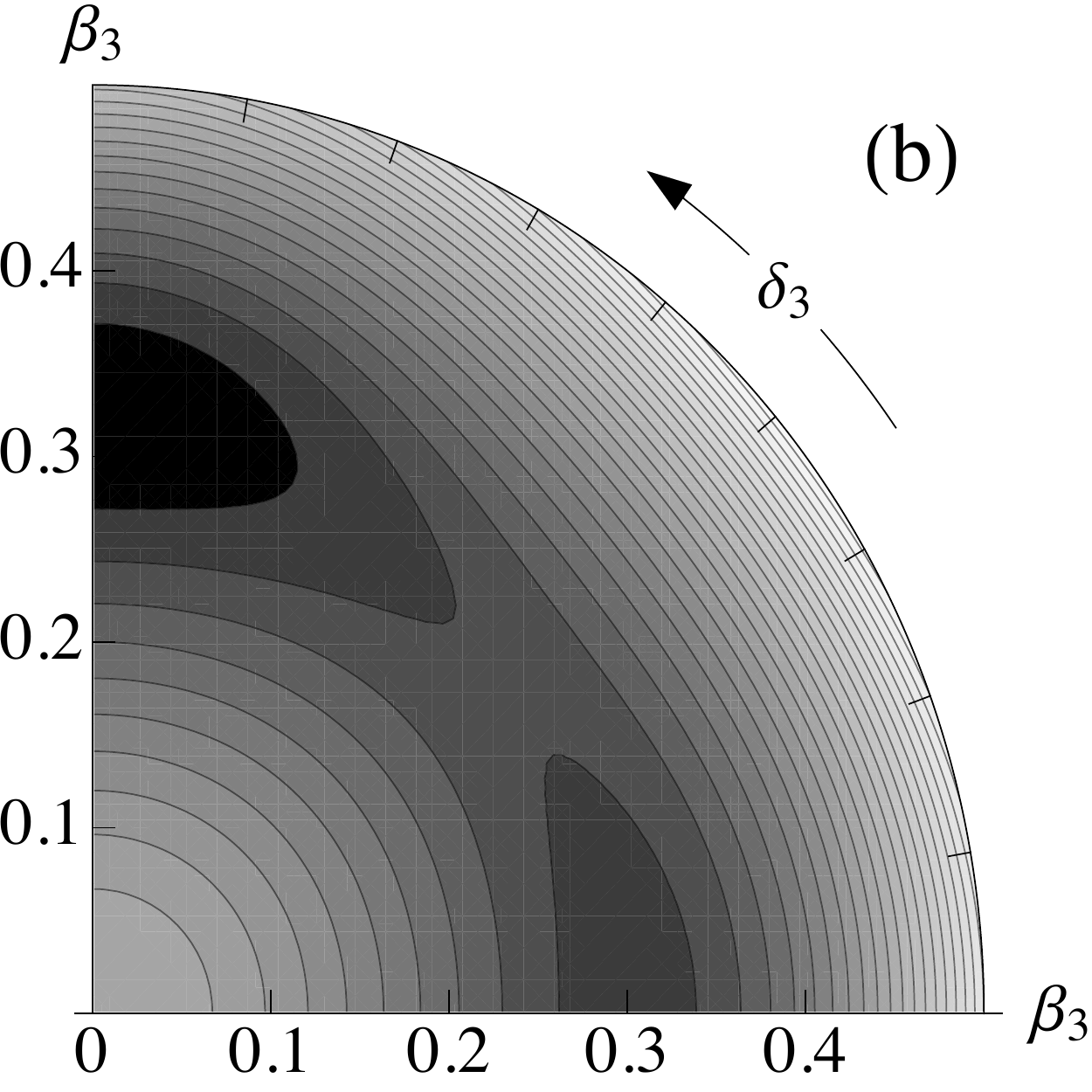}
\includegraphics[width=4cm]{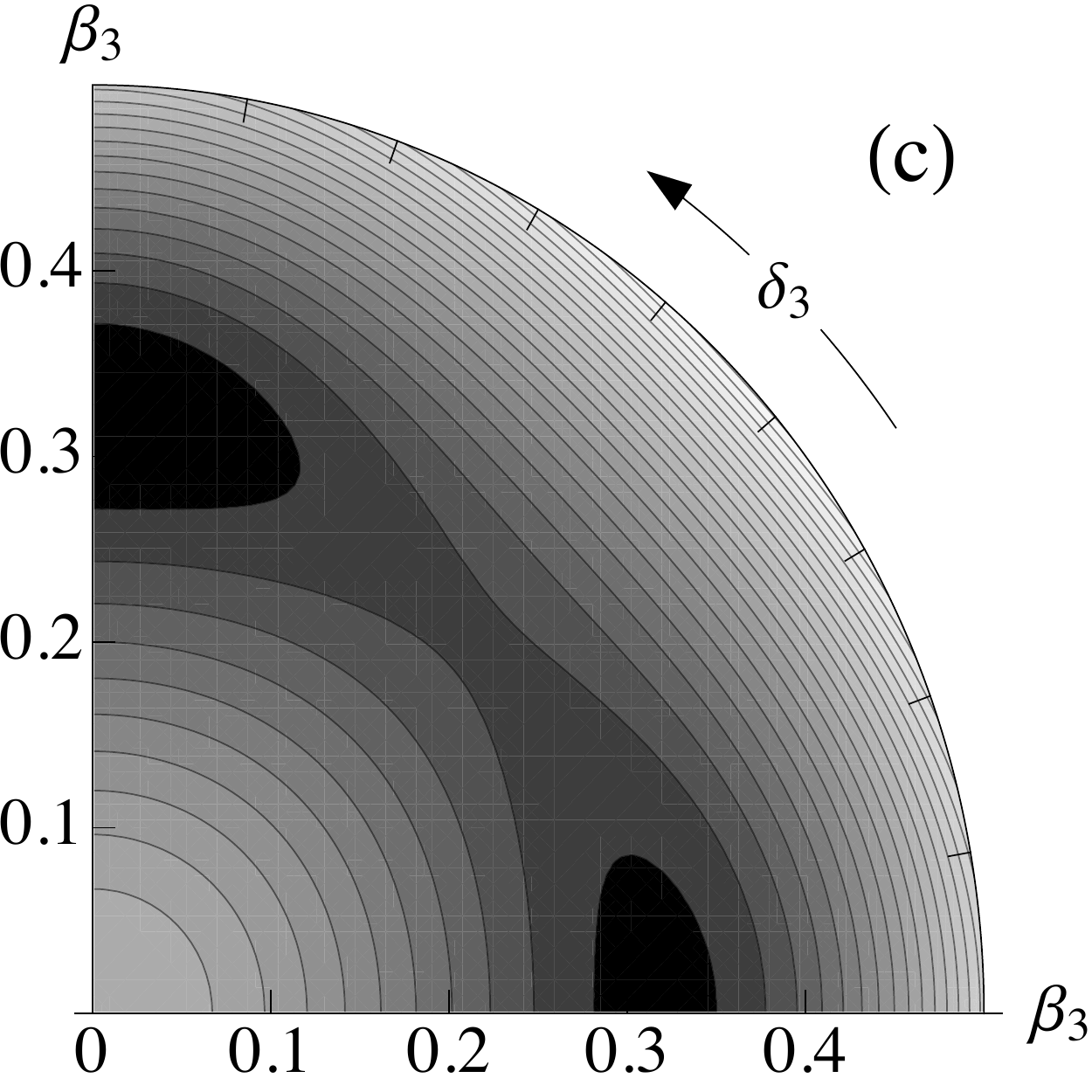}
\caption{Three energy surfaces $E(\beta_3,\delta_3,\vartheta_3^*,\varphi_3^*)$
obtained in the classical limit
of two different Hamiltonians of the \mbox{$sf$-IBM} for $N=6$ bosons.
The values of $\vartheta_3$ and $\varphi_3$ are fixed
and the dependence on $\beta_3>0$ and $0\leq\delta_3\leq{\frac12}\pi$ is shown.
Black corresponds to low energies
and the lines indicate changes by 10~keV.
(a)  The ${\rm U}_f(7)$--${\rm SO}_{sf}(8)$ transitional Hamiltonian is taken
with the parameters given in the caption of Fig.~\ref{f_u7so8}.
(b) and (c) The Hamiltonian of (a) is modified
by taking a repulsive interaction $v_{ffff}^2=500$~keV.
The energy surface is shown 
for (b) $\vartheta_3^*={\frac12}\pi$ and $\varphi_3^*=0$,
and for (c) $\vartheta_3^*={\frac12}\pi$ and $\varphi_3^*={\frac14}\pi$.}
\label{f_b3d3}
\end{figure}
The parameters quoted in the caption of Fig.~\ref{f_b3d3}
satisfy the conditions~(\ref{e_conlat12})
and, as a result, the energy surface in the classical limit
of the corresponding Hamiltonian
displays an octupole-deformed minimum.
According to the preceding discussion,
the energy surface is independent of $\delta_3$
unless the matrix element $v_{ffff}^2$ is made repulsive,
in which case an isolated tetrahedral minimum develops.
This is indeed confirmed by the surfaces shown in Fig.~\ref{f_b3d3},
obtained by taking the classical limit of two different Hamiltonians of the \mbox{$sf$-IBM}.
For display purposes the values of $\vartheta_3$ and $\varphi_3$ are fixed
and the dependence on $\beta_3>0$ and $0\leq\delta_3\leq{\frac12}\pi$ is shown.
The classical limit of the symmetry Hamiltonian~(\ref{e_hamlat}),
for which $v_{ffff}^2=2a_f+2b_f-18c_f$,
displays an octupole-deformed minimum at $\beta_3^*\approx0.32$
and no dependence of $\delta_3$, as shown in Fig.~\ref{f_b3d3}(a).
The change to $v_{ffff}^2=500$~keV
introduces a minimum with tetrahedral symmetry ($\delta_3^*={\frac12}\pi$)
as shown in Fig.~\ref{f_b3d3}(b) for $\vartheta_3^*={\frac12}\pi$ and $\varphi_3^*=0$,
and in Fig.~\ref{f_b3d3}(c) for $\vartheta_3^*={\frac12}\pi$ and $\varphi_3^*={\frac14}\pi$.
The latter energy surfaces display a second minimum
with axially symmetric octupole deformation ($\delta_3^*=0$).

\begin{figure}
\centering
\includegraphics[width=3cm]{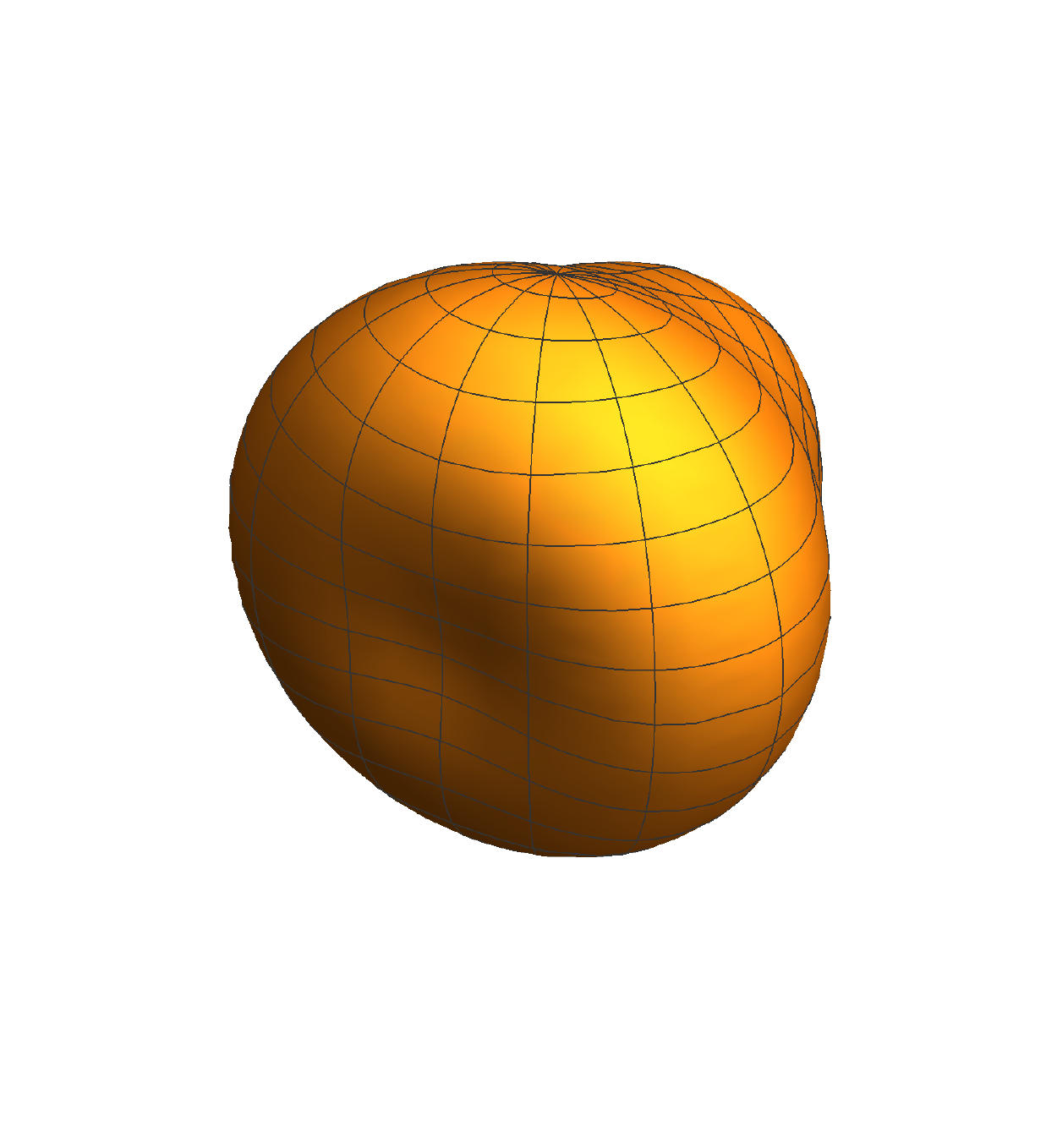}
\includegraphics[width=7cm]{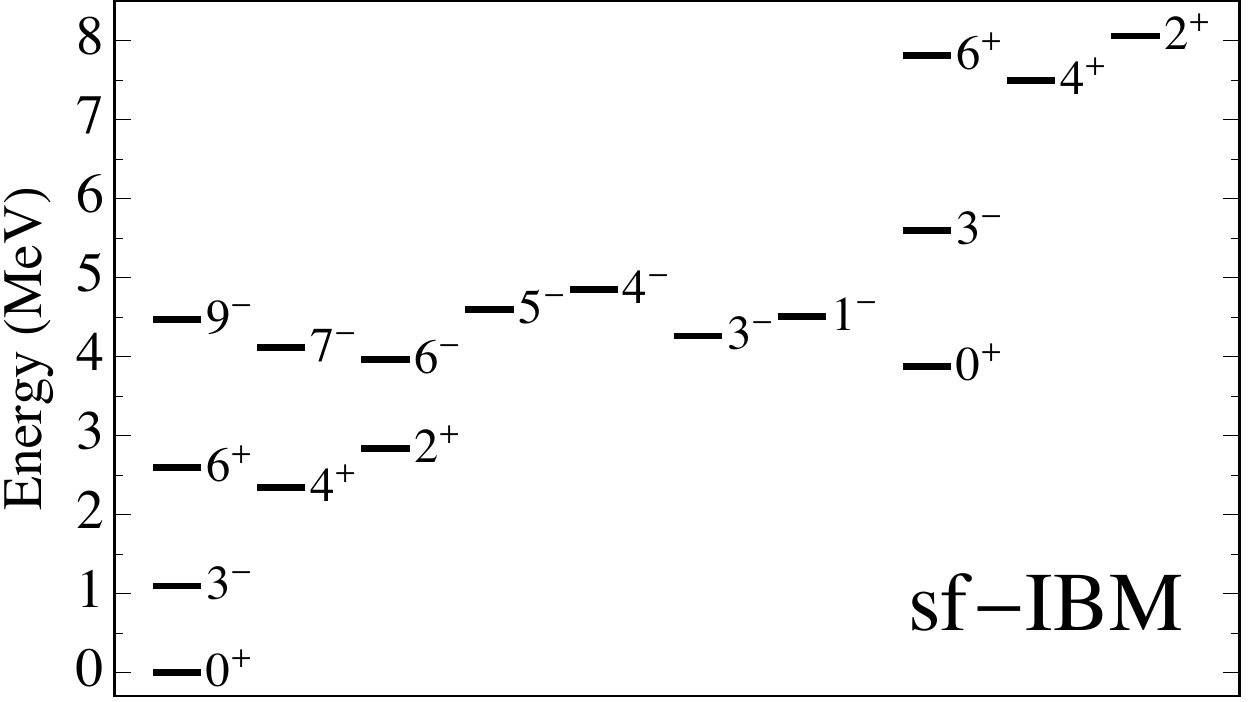}
\includegraphics[width=3cm]{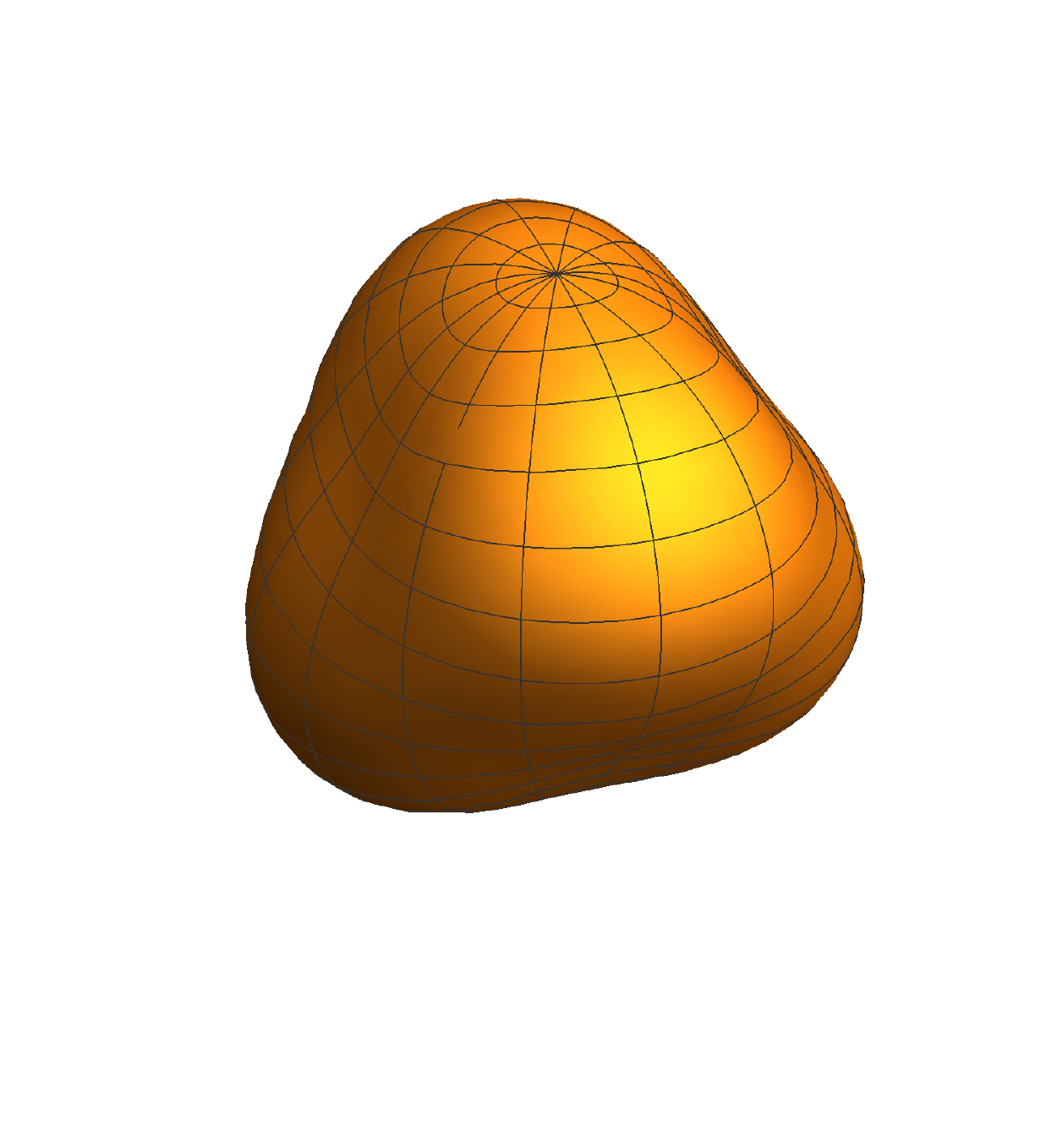}
\caption{Energy spectrum of a general Hamiltonian of the \mbox{$sf$-IBM} for $N=6$ bosons.
The same Hamiltonian is taken as in Fig.~\ref{f_u7so8}
but one $f$-boson two-body matrix element is modified to $v_{ffff}^2=500$~keV.
On the left- and right-hand sides are shown
the shapes at the minima in the energy surface
obtained in the classical limit of this Hamiltonian.
The shape on the left is axially symmetric, octupole deformed
while the shape on the right has tetrahedral symmetry.}
\label{f_u8}
\end{figure}
Although this proves that shapes with tetrahedral symmetry
may occur with a reasonable parameterization in the \mbox{$sf$-IBM},
it is to be expected that the minimum is rather shallow
as it occurs as a result of fine-tuning of little-known $f$-boson interactions.
Even with a value as large as $v_{ffff}^2=500$~keV,
the tetrahedral ($\delta_3^*={\frac12}\pi$) and the axially symmetric ($\delta_3^*=0$) minima
are separated by a barrier of only $\sim20$~keV.
As a result, only minute observable effects can be expected.
This is illustrated in Fig.~\ref{f_u8},
which shows the spectrum of the ${\rm U}_f(8)$--${\rm SO}_{sf}(8)$ transitional Hamiltonian
with the modified $v_{ffff}^2$ matrix element.
Not much difference from the spectrum shown in Fig.~\ref{f_u7so8} can be seen.

\section{Conclusions}
\label{s_conc}
Two dynamical symmetries of the \mbox{$sf$-IBM} have been established:
the ${\rm U}_f(7)$ limit with octupole vibrational characteristics
and the ${\rm SO}_{sf}(8)$ limit
where $s$- and $f$-boson states
are mixed through an $sf$-pairing interaction,
which, if strong enough, drives the system towards a permanent octupole deformation.
This picture is confirmed by a catastrophe analysis of the energy surface
obtained in the classical limit of a Hamiltonian transitional between the two limits,
indicating that an octupole-deformed minimum
can be obtained with reasonable single-boson energies.
However, this minimum is always $\delta_3$ independent
and shapes ranging from pear-like to tetrahedral are degenerate in energy.
An isolated minimum with tetrahedral symmetry can be obtained
by modifying two-body interactions between the $f$ bosons
to the transitional symmetry Hamiltonian.
It is separated from another minimum with axial symmetry by a low-energy barrier,
even for fairly strong interactions between the $f$ bosons.

There are striking similarities
between the search for tetrahedral shapes presented in this paper
and the corresponding search for octahedral shapes reported in I and II.
In both cases it is found that no isolated minimum with a higher-rank discrete symmetry
is possible for a symmetry Hamiltonian of U(8) or U(15)
but that a degenerate minimum occurs
in the ${\rm SO}_{sf}(8)$ or ${\rm SO}_{sg}(10)$ limits
of $sf$- or $sg$-pairing, respectively.
An isolated minimum with tetrahedral or octahedral symmetry
can be obtained through a modification of the two-body interaction between the relevant bosons.
However, the minima thus constructed are rather shallow,
even for large repulsive matrix elements between the $f$ or $g$ bosons,
and their effects on spectroscopic properties are expected to be minute.

With this series of papers the role of higher-rank discrete symmetries
in the context of algebraic nuclear models is clarified
and a well-defined procedure is established to find out
whether a given Hamiltonian of a particular version of the interacting boson model
displays in its classical limit a minimum with a tetrahedral or octahedral shape.
This enables the study of observable consequences
of higher-rank discrete symmetries in the framework of algebraic models.

The limitations of this series of papers should nevertheless be recognized
because the present analysis is restricted to Hamiltonians with up to two-body terms.
It is possible that, just as triaxial shapes require higher-order interactions in the \mbox{$sd$-IBM},
shapes with a higher-rank discrete symmetry can be isolated with a high barrier
by introducing higher-order interactions in \mbox{$sdg$-IBM} and \mbox{$sf$-IBM}.
Also, as mentioned in the introduction of this paper,
the analysis of the tetrahedral case so far has been limited to \mbox{$sf$-IBM}
and should be carried out in the more general \mbox{$spdf$-IBM}.
What can be concluded from the examples reported in this series of papers
is that, unless such more complicated Hamiltonians are adopted,
it will be difficult to identify clear effects of higher-rank discrete symmetries in nuclei.



\begin{thebibliography}{99}
\bibitem{note1}
The $\mu=0$ term is treated here separately
to avoid the double counting in Eq.~(12) of I.
\end{thebibliography}
\end{document}